\documentclass[a4paper,english,8pt]{article}
\usepackage{graphicx}
\usepackage{epsf}
\usepackage{amsmath}
\usepackage{amssymb}
\usepackage[T2A]{fontenc}
\usepackage[cp1251]{inputenc}
\usepackage{babel}
\usepackage[labelsep=period,hang]{caption} 
\newcommand{\NMS}[1]{\mbox{$#1\,M_{\textstyle\odot}$}}
\newcommand{\xmn}[2]{\mbox{$#1\!\times\! 10^{#2}\,$}}
\newcommand{\isn}[2]{\mbox{$^{#2}${#1}}}
\newcommand{\avr}[1]{\mbox{$\langle #1\rangle$}}
\newcommand{\indis}[1]{{\mbox{\scriptsize #1}}}

\newcommand{\nucmu}{\mbox{$m_{\mathrm u}$}}

\newcommand{\D}{\widehat{\mathrm D}}
\newcommand{\B}{\widehat{\mathrm B}}
\newcommand{\feq}{f_{\mathrm{e}}}

\newcommand{\lamb}{\lambda_\mathrm{m}}
\newcommand{\lambn}{\lambda_{\mathrm{m}\nu}}
\newcommand{\lamban}{\lambda_{\mathrm{m}\overline{\nu}}}

\newcommand{\gT}{g_T}
\newcommand{\gpsi}{g_{\psi}}
\newcommand{\epsn}{\mbox{$\omega_{\nu}$}}

\newcommand{\KNU}{\mbox{$K_{\nu}$}}
\newcommand{\KNUOE}{\mbox{$K_{\mathrm{e}0}$}}
\newcommand{\SNU}{\mbox{$S_{\nu}$}}

\newcommand{\UNU}{\mbox{$U_{\nu}$}}
\newcommand{\UNUOE}{\mbox{$U_{\mathrm{e}0}$}}
\newcommand{\BINU}{\mbox{${\rm {\bf I}}_{\nu}$}}
\newcommand{\BINUEO}{\mbox{${\rm {\bf I}}_0$}}
\newcommand{\INU}{\mbox{$I_{\nu}$}}
\newcommand{\INUA}{\mbox{$I_{\overline{\nu}}$}}
\newcommand{\INUE}{\mbox{$I_{\nu\mathrm{e}}$}}
\newcommand{\IE}{\mbox{$I_{\mathrm{e}}$}}
\newcommand{\WID}[2]{#1_{\mathrm{#2}}}

\begin{document}

\title{Principal Physical Effects in Collapsing Stellar Cores}

\author{D.K.~Nadyozhin and A.V.~Yudin%
        \thanks{E-mail: {nadezhin@itep.ru}, {yudin@itep.ru}}\\
{\it  Institute for Theoretical and Experimental Physics, Moscow, Russia}\\
{\it        SRC Kurchatov Institute, Moscow, Russia}}

\date{}

\maketitle
\abstract{
 Principally important for the description the physical processes
 in the collapsing stellar cores topics are surveyed.
 They are: the neutrino heat conduction theory, equation of
 state under the conditions of nuclear statistical equilibrium
 and possible phase transitions in dense, subnuclear and
 nuclear environment.
 }

\section*{The Neutrino Heat Conduction Theory
           in Collapsing Stellar Cores}

 The neutrino transport in collapsing stellar core:
   \begin{itemize}
\item Determines dynamic and thermodynamic properties of implosion
  such as density, temperature and lepton number at the moment of bounce;
\item  Of crucial importance for possible conversion of stalled shock into
   an outgoing blast wave.
\item  The major part of neutrino flux is radiated from  neutrinosphere
   in the regime of neutrino opacity.
   The flux is controlled by neutrino diffusion
   in the neutrino-opaque core and can be well described
   by the Neutrino Heat Conduction (NHC) theory.
   \end{itemize}

\subsubsection*{
 The equations of neutrino hydrodynamics}

In the spherically symmetric case, the standard
expression for the energy–momentum tensor \cite{LandauLifshitz1945}
of a system that consists of
matter and radiation leads to the following equations
of neutrino hydrodynamics:
\begin{eqnarray}
 & & \frac{du}{dt}
 = -\frac{1}{\rho}\frac{\partial P}{\partial r}\: - \frac{Gm}{r^{2}}
 -\frac{1}{\rho}\mathcal{M}_{\nu}\, ,\label{ngd1} \\
 & & \frac{dE}{dt}\: +\: P \frac{d}{dt}\left(\frac{1}{\rho }
\right) =
 \frac{1}{\rho}\left(u\,\mathcal{M}_{\nu}-\mathcal{E}_{\nu}\right),
 \label{ngd2}\\
 & & \frac{\partial \rho }{\partial t}\: +\: \frac{1}{r^{2}}
\frac{\partial }{\partial r}\left(r^{2}\rho u\right)\:  = \: 0 \; ,
\label{ngd3}\\
 & & \frac{\partial m}{\partial r} = 4\pi r^2\,\rho\, ,
\label{ngd4}
\end{eqnarray}
where $u$, $\rho$, $P$, and $E$ are the velocity, density, pressure,
and specific energy of the matter, respectively.
Equations (\ref{ngd4}) are written in the laboratory (rest)
frame with Newtonian gravity. The terms of order
$(u/c)^2$ and $(a_s/c)^2$ ($c$ and $a_s$ are the speeds of light
and sound, respectively) were discarded.

 The terms $\mathcal{M}_{\nu}$ and $\mathcal{E}_{\nu}$
describe the momentum and energy exchange between
matter and neutrino radiation and are defined by
\begin{eqnarray}
\mathcal{M}_{\nu} & = &\frac{1}{c^{2}}
\frac{\partial S_{\nu}}{\partial t} +
\frac{\partial K_{\nu}}{\partial r} + \frac{1}{r}
\left(3\KNU-\UNU\right),\label{Momrate} \\
\mathcal{E}_{\nu} & = & \frac{\partial\UNU}{\partial t}
 + \frac{1}{r^{2}}
\frac{\partial}{\partial r}\left(r^{2}\SNU\right) ,
\label{Enrate}
\end{eqnarray}
where \KNU\ , \SNU\ , and \UNU\ are the angular moments of
the total (energy-integrated) neutrino radiation intensity
\BINU\ :
\begin{eqnarray}
\KNU =\,\KNU(r,t) & = & \frac{2\pi}{c}\int_{-1}^{1}
\mu^{2}\,\BINU\, d\mu \, ,\label{KMom}\\
\SNU =\,\SNU(r,t) & = & 2\pi\int_{-1}^{1}
\mu\,\BINU\, d\mu\, ,\label{SMom}\\
\UNU =\,\UNU(r,t) & = & \frac{2\pi}{c}\int_{-1}^{1}
\BINU\, d\mu\, ,\label{UMom}
\end{eqnarray}
where $\mu$ is the cosine of the angle between the
neutrino propagation direction and the radius vector.
Physically, \KNU\ and \SNU\ define the neutrino-transferred
momentum and energy fluxes, while \UNU\ is the neutrino
energy density. The total intensity \BINU\ is an
integral of the spectral intensity \INU\  over the neutrino
energy \epsn\:
\begin{equation}
\BINU = \BINU(\mu,r,t) = \int_{0}^{\infty}\!\INU(\epsn,\mu,r,t)
 \, d\epsn\, .\label{Itot}
\end{equation}

The spectral intensity \INU\ is described by a transport
equation that should include both true neutrino
absorption and scattering.

\subsubsection*{
    The neutrino transfer equation with scattering}

In the general case, the neutrino transport equation
in the laboratory frame reads
\begin{equation}
 \D\INU = -\frac{\INU - \INUE}{\widetilde{l}_{\nu}} + \B(\INU,\INU^\prime)\, ,
 \label{Transeq}
\end{equation}
\begin{equation}
 \D \equiv \frac{1}{c}\frac{\partial }{\partial t} +
 \mu \frac{\partial}{\partial r} +
\frac{1-\mu ^{2}}{r}\frac{\partial }{\partial \mu}\, .
\label{Dhatop}
\end{equation}
where $\D$ denotes a linear differential operator that describes
the neutrino redistribution in space and
\INUE\ is the equilibrium intensity,
$\widetilde{l}_{\nu}$ is the neutrino
mean free path with respect to the true absorption
with allowance made for the induced absorption.

The neutrino scattering is described by a nonlinear
integral operator $\B\,$ whose explicit form is considered
below (the nonlinearity appears due to the Pauli
exclusion principle). So far it is suffice for us to know
that the intensity enters $\B\,$ both directly (\INU) and
under the integral sign ($\INU^\prime$).
We can now ascertain how the quantities $\mathcal{E}_{\nu}$ and
$\mathcal{M}_{\nu}$ introduced in the previous section are expressed
in terms of the intensity. First, we multiply
Eq. (\ref{Transeq}) by $2\pi$ and integrate it over $\mu$ and \epsn .
Then we repeat the same procedure after multiplication
by $2\pi\mu/c\,$.
As a result, we obtain using Eqs. (\ref{KMom})--(\ref{UMom})
\begin{equation}
 \mathcal{E}_{\nu}= 2\pi\!\int_{0}^{\infty}\!\!\!\int_{-1}^{1}
\!\!\left[-\frac{\INU-\INUE}{\widetilde{l}_{\nu}}+ \B(\INU,\INU^\prime)
\right]d\mu\, d\epsn\, ,
\label{Entran}
\end{equation}
\begin{equation}
 \mathcal{M}_{\nu}=
\frac{2\pi}{c}\!\int_{0}^{\infty}\!\!\!\int_{-1}^{1}\mu
\left[-\frac{\INU-\INUE}{\widetilde{l}_{\nu}}+ \B(\INU,\INU^\prime)
\right]d\mu\, d\epsn\, .
\label{Momtran}
\end{equation}
\subsubsection*{The method of successive approximations}

We will seek a solution to the transport equation by
an expansion in terms of a small parameter—the neutrino
mean free path $l_\nu$ with respect to the absorption
and scattering (the effective mean free path relative to
the scattering is contained in the operator $\B(\INU,\INU^\prime)$).
Below, for simplicity, we will no longer write the subscript $\nu$.
The first approximation $I_1$ for the intensity $I$
can be obtained by substituting its equilibrium value
$I_{\mathrm{e}}$ into the left-hand side of Eq. (\ref{Transeq})
 — into the operator $\D\, I\,$:
 \begin{equation}\label{AI1}
  \D I_{\mathrm{e}} = -\frac{I_1-\IE}{\widetilde{l}}+ \B(I_1,I_1^\prime)
    \, .
 \end{equation}
Solving this integral equation for $I_1$ and substituting
the derived $I_1$ into the operator $\D I$ yields an integral
equation for the second approximation $I_2\,$:
 \begin{equation}\label{AI2}
    \D I_1 = -\frac{I_2-\IE}{\widetilde{l}}+ \B(I_2,I_2^\prime)\, .
 \end{equation}
The NHC equations can be derived
when using the right-hand side of Eq. (\ref{AI2}) in the
integrands of Eqs. (\ref{Entran}) and (\ref{Momtran}).
However, there is no need to solve Eq. (\ref{AI2}),
since its right-hand side is just equal to $\D I_1\,$.
Hence, we can write
\begin{equation}
 \mathcal{E}= 2\pi\!\int_{0}^{\infty}\!\!\!\int_{-1}^{1}\!
 \D I_1\, d\mu\, d\omega\, ,
\label{Entran1}
\end{equation}
\begin{equation}
 \mathcal{M}=
\frac{2\pi}{c}\!\int_{0}^{\infty}\!\!\!\int_{-1}^{1}\mu
\D I_1\, d\mu\, d\omega\, .
\label{Momtran1}
\end{equation}
A key point is the solution of Eq. (\ref{AI1}) for $I_1$.
Let us represent $I_1$ as
 \begin{equation}\label{I1PDEL}
    I_1 = \IE + \delta I_1\, ,
 \end{equation}
where $\delta I_1$ is a small correction to the equilibrium
intensity $\delta I_1$ proportional to the neutrino mean free
path.

To calculate the integrals on the right-hand sides
of Eqs. (\ref{Entran1}) and (\ref{Momtran1}),
we now must pass to a comoving frame
where the matter is at rest $(u=0)$ and where the
mean free paths with respect to the neutrino absorption
and scattering are defined explicitly. This makes it
possible to perform the integration over $\mu$ analytically.

\subsubsection*{Passing to a comoving frame}

We will denote the quantities pertaining to the
comoving frame by the subscript 0. If the terms of
order $(u/c)^2$ are disregarded, then the quantities in
the laboratory and comoving frames are related by
\begin{equation}\label{Lorenz1}
\begin{aligned}
 t &=t_0\, ,& r &=r_0\, , \\
 \omega &=\omega_0/L\, ,& \mu &=\mu_0+(1-\mu_0^2)u/c\, ,
\end{aligned}
\end{equation}
where the Lorentz factor $L$ in our approximation is
\begin{equation}\label{factorL}
 L = 1-\mu u/c = 1-\mu_0 u/c\, .
\end{equation}
In addition, the following transformations for the intensity,
mean free path, and scattering operator follow
from the Lorentz invariance of the transport equation
\cite{Thomas1930,ImshMoroz1981}, 
\begin{equation}\label{Lorenz2}
\begin{aligned}
 I &=I_0/L^3\, ,& l &=l_0/L\, , & \B & =\B_0/L^2\, .
\end{aligned}
\end{equation}

The equilibrium intensity in the comoving frame
does not depend on the neutrino propagation direction:
\begin{equation}\label{Ieq0}
 I_{\mathrm{e0}}\: =\:\frac{\omega_0^3}{c^2h^3}
\frac{1}{1 + \exp\left(\frac{\omega_0}{kT}- \psi\right)}\, ,
\end{equation}
where $h$ is the Planck constant, $k$ is the Boltzmann
constant, and $\psi$ is the neutrino chemical potential in
units of $kT$. However, $I_{\mathrm e}$ is anisotropic
in the laboratory frame (at fixed neutrino energy $\omega$):
\begin{equation}\label{Ieq}
 I_{\mathrm{e}}\: =\:\frac{\omega^3}{c^2h^3}
\frac{1}{1 + \exp\left(\frac{L\omega}{kT}- \psi\right)}\,
=\left(\frac{\omega}{\omega_0}\right)^3 I_{\mathrm{e}0}.
\end{equation}

It is convenient to introduce a dimensionless neutrino
distribution function $f$ that describes the degree
of phase-space filling and that is an invariant of the
Lorentz transformations:
 \begin{equation}\label{finvar}
 f\, =\,\frac{c^2h^3}{\omega^3}I\, =\,\frac{c^2h^3}{\omega_0^3}I_0\, .
 \end{equation}
The equilibrium value of $f$ is
 \begin{equation}\label{finvareq}
 f_{\mathrm{e}}\, =\,\frac{1}{1 + \exp\left(\frac{\omega_0}{kT}-\psi\right)}
 \, =\,\frac{1}{1 + \exp\left(\frac{L\omega}{kT}- \psi\right)}\, .
 \end{equation}
The equilibrium energy density \UNUOE\ and pressure \KNUOE\
in the comoving frame are
\begin{eqnarray}
\UNUOE\ =\, 3\KNUOE & = &\frac{2\pi }{c}\int _{-1}^{1}\BINUEO\ d\mu_0 =
 \frac{4\pi}{c^3h^3}\int_{0}^{\infty}\!\omega_0^3\, f_{\mathrm{e}}\, d\omega_0
\nonumber\\
& = & \frac{15aT^{4}}{2\pi ^{4}}\: F_{3}(\psi)\, ,
\quad \left(a\equiv\frac{\pi ^{2}k^{4}}{15\hbar^{3}c^{3}}\right)\, ,
\label{UK0E}
\end{eqnarray}
where $F_3$ is the Fermi–Dirac function of index 3.
The equilibrium neutrino number density $n_{\mathrm{e}0}$ can be
expressed in terms of the Fermi–Dirac function of
index 2:
\begin{equation}\label{NNUE}
n_{\mathrm{e}0}\, =\,
\frac{4\pi}{c^3h^3}\int_{0}^{\infty}\!\omega_0^2\, f_{\mathrm{e}}\, d\omega_0
=\frac{15aT^3}{2\pi ^4 k}\: F_2(\psi)\, .
\end{equation}
Similar expressions for antineutrinos can be derived
by substituting $\psi$ for $-\psi$.

Applying the operator $\D$ to both sides of Eq. (\ref{I1PDEL}),
we obtain
 \begin{equation}\label{D1PDEL}
    \D I_1=\D I_{\mathrm{e}}+\D\delta I_1\, .
 \end{equation}
Now, everything is ready for the integration of $\D I_1$ in
Eqs. (\ref{Entran1}) and (\ref{Momtran1}).
In accordance with Eqs. (\ref{Lorenz1})--(\ref{Lorenz2}),
we pass from the integration variables  $(\mu,\omega)$
to $(\mu_0,\omega_0)$, given that
$d\omega = d\omega_0 /L$ and $d\mu =L^2 d\mu_0$
and that the limits of integration over $\mu$ and $\omega$ do
not change. In addition, we will disregard not only
the small quantities of order $(u/c)^2$, but also the
quantities of order $(u/c)l\frac{\partial}{\partial r}$,
which describe the neutrino
radiation viscosity. Since $\delta I_1$ has the order of
smallness $l\frac{\partial}{\partial r}$,
the differences between the laboratory
and comoving frames with the order of smallness
u/c should be disregarded in this approximation both
when calculating $\delta I_1$ and when integrating $\D\delta I_1$.

It will be clear from the subsequent analysis that
$\delta I_1$ is proportional to $\mu_0$ in the comoving frame
(see also p. 71 in paper \cite{ImshMoroz1981})
\begin{equation}\label{DImu}
 \delta I_1\, =\,\mu_0\, G_1(\omega_0)\, ,
\end{equation}
where $G_1$ no longer depends on the neutrino propagation direction.
Performing the integration
in Eqs. (\ref{Entran1}) and (\ref{Momtran1}) by
taking into account all of these remarks, we ultimately
obtain
\begin{eqnarray}
 \mathcal{E} & = &
\frac{\partial\UNUOE}{\partial t}
 + \frac{1}{r^{2}}\frac{\partial}{\partial r}\left[r^{2}\left(
\frac{4}{3}u\UNUOE + \frac{4\pi }{3}
\int_{0}^{\infty }\! G_1\, d\omega_0\right)\right]\, ,\label{Entranf}\\
 \mathcal{M} & = & \frac{\partial \KNUOE\ }{\partial r}\, .\label{Momtranf}
\end{eqnarray}
In the absence of scattering,
 \begin{equation}\label{Gab}
  G_1(\omega_0)\, =\, -\widetilde{l}_0\frac{\partial I_{\mathrm{e}0}}{\partial r}\,
  = \, -\widetilde{l}_0\left(\frac{\partial I_{\mathrm{e}0}}{\partial T}
  \frac{\partial T}{\partial r}
  +\frac{\partial I_{\mathrm{e}0}}{\partial\psi}\frac{\partial\psi}
  {\partial r}\right).
 \end{equation}
It is this value of $G_1(\omega_0)$ that was used in
paper \cite{ImshNad1972}.

\subsubsection*{The scattering operator}

Since all quantities in this section are considered in the
comoving frame, we omit the subscript 0.

In the comoving frame, the scattering operator
can be written as \cite{BludmanVRiper1978,ImshNad1979,Nad1994,Cernohorsky1994}:
 \begin{eqnarray}
 \B (I,I^\prime)=\!\!\int\!\!\!\int\!\!\left[\frac{\omega}{\omega^\prime}
 R\left(\omega^\prime,\omega,\eta\right)I^\prime(1{-}f)-
 R\left(\omega,\omega^\prime ,\eta\right)I(1{-}f^\prime)\right]
 d\Omega^\prime\, d\omega^\prime\nonumber\\
 =\frac{\omega}{c^2h^3}\!\!\int\!\!\!\int\!\!\left[\omega^{\prime 2}
 R\left(\omega^\prime,\omega,\eta\right)f^\prime(1{-}f)-
 \omega^2 R\left(\omega,\omega^\prime ,\eta\right)f(1{-}f^\prime)\right]
 d\Omega^\prime\, d\omega^\prime ,\label{Bhat}
 \end{eqnarray}
where the scattering kernel $R\left(\omega,\omega^\prime ,\eta\right)$
depends on three arguments: the neutrino energies before and
after the scattering (the first and second arguments,
respectively) and the cosine $\eta$ of the angle between the
neutrino momentum vectors before and after the scattering.
This dependence is determined by the microscopic
properties of an elementary scattering event.

The first term in the integrand in Eq. (\ref{Bhat}) takes into
account the contribution from the neutrinos with energy
$\omega'$ before their scattering and those scattered
in the direction м with energy щ, while the second
term describes the “knocking-out” of neutrinos with
energy $\omega$ from the beam as a result of their scattering
in an arbitrary direction with a change in energy. The
integration in Eq. (\ref{Bhat}) is over the energy and solid angle
$(d\Omega^\prime =\sin\theta^\prime d\theta^\prime d\phi^\prime =
 d\mu^\prime d\phi^\prime)$
of the neutrinos before
and after their scattering in the first and second terms
of the integrand, respectively.

In total thermodynamic equilibrium,
$I=I_{\mathrm{e}}$ and $I^\prime =I_{\mathrm{e}}^\prime$,
the value of $\B(I,I^\prime)$ should become zero: the
number of neutrinos escaping from the beam as a result
of their scattering should be exactly equal to their
number scattered in the beam direction (the principle
of detailed balancing!).
The condition $\B(I_\mathrm{e},I_\mathrm{e}^\prime)=0$
yields the following property of the kernel $R$ with
respect to the interchange of its arguments $\omega$
and $\omega^\prime$:
  \begin{equation}\label{Rexch}
  R\left(\omega^\prime,\omega,\eta\right)\, =\,
  \left(\frac{\omega}{\omega^\prime}\right)^2
  \mathrm{e}^\frac{\omega^\prime -\omega}{kT}
  R\left(\omega,\omega^\prime,\eta\right)\, .
  \end{equation}
Given this expression, the scattering operator takes
the form
  \begin{equation}\label{Bfin}
  \B (I,I^\prime)=\frac{\omega^3}{c^2 h^3}
  \!\!\int\!\!\!\int\!\!\left[\mathrm{e}^\frac{\omega^\prime -\omega}{kT}
  f^\prime\left(1{-}f\right) - f\left(1{-}f^\prime\right)\right]\!\!
   R\left(\omega,\omega^\prime,\eta\right)
    d\Omega^{\prime}d\omega^{\prime}.
  \end{equation}
When integrating over $\Omega^\prime$, it should be kept in mind
that
  \begin{equation}\label{etamu}
  \eta\, =\, \mu\mu^\prime +\sqrt{1-\mu^2}\sqrt{1-\mu^{\prime 2}}
  \cos\left(\phi -\phi^\prime\right)\, .
  \end{equation}
The mean free path $l_\mathrm{s}$ with respect to the scattering
is related to $R$ by
  \begin{equation}\label{indicat}
   l_\mathrm{s}^{-1}\, =\int\!\!\!\int\!\!
   R\left(\omega,\omega^\prime,\eta\right)\, d\Omega^\prime\, d\omega^\prime .
  \end{equation}
Consequently, $l_\mathrm{s}R\left(\omega,\omega^\prime,\eta\right)$
is the scattering indicatrix for a neutrino with energy
$\omega$ scattered with a change in energy by $\omega^\prime$
through the angle $\arccos\eta$
with respect to the direction of its initial propagation.

If the scattering is coherent (conservative), i.e.,
occurs without any change in energy $(\omega =\omega')$, as,
for example, in the case of neutrino scattering by
nucleons and atomic nuclei, then $R$ can be written as
  \begin{equation}\label{Rdelta}
  R\left(\omega,\omega^\prime,\eta\right)=R_\mathrm{cs}(\omega,\eta)
  \,\delta(\omega-\omega^\prime)\, ,
  \end{equation}
where $\delta$ is the Dirac delta function. In this case, the
expression for $\B$ is simplified significantly:
  \begin{equation}\label{Bcoh}
  \B (I,I^\prime)=\,\frac{\omega^3}{c^2 h^3}
  \int\!\!\left(f^\prime -f\right)
   R_\mathrm{cs}(\omega,\eta)\,d\Omega^\prime\, .
  \end{equation}
The factors $(1-f)$ and $(1-f^\prime)$ in the in and out
beams, respectively, that describe the Pauli exclusion
principle cancel each other out. Hence, for coherent
scattering, the Pauli exclusion principle for neutrinos
may be disregarded. Here, there is a close analogy
with the compensation of the induced emission in
the case of Thomson photon scattering \cite{DAFK1959}.

\subsubsection*{The form of the integral equation}

Let us now turn to the transformation of the integral
equation (\ref{AI1}), which, according to the aforesaid,
should be considered in the comoving frame.
Substituting $I_1$ from Eq. (\ref{I1PDEL}) into (\ref{AI1}),
we obtain for $\delta I_1$
  \begin{equation}\label{AI1del}
   \delta I_1\, =\, - \widetilde{l}\:\D I_{\mathrm{e}}
   + \widetilde{l}\:\B(I_{\mathrm{e}}+\delta I_1,I_{\mathrm{e}}^\prime
   +\delta I_1^\prime)\, .
  \end{equation}
Discarding the small terms of order $\delta I_1^2$
and taking into account $\B(I_{\mathrm{e}},I_{\mathrm{e}}^\prime) =0$,
the scattering operator takes the form
  \begin{equation}\label{Blinear}
   \B(I_{\mathrm{e}}+\delta I_1,I_{\mathrm{e}}^\prime +\delta I_1^\prime)
   =\int\!\!\!\int\!\!\left(
   \frac{I_\mathrm{e}}{I_\mathrm{e}^\prime}\,\delta I_1^\prime -
   \frac{1-f_\mathrm{e}^\prime}{1-f_\mathrm{e}}\,\delta I_1\right)
   R\left(\omega,\omega^\prime,\eta\right)\, d\Omega^\prime\, d\omega^\prime .
  \end{equation}
The time derivative in the operator $\D$ in Eq. (\ref{AI1del}) leads
to terms of the order of smallness
$\widetilde{l}\:\frac{1}{c}\frac{\partial}{\partial t}\sim
   \widetilde{l}\:\frac{u}{c}\frac{\partial}{\partial r}$,
which we will also disregard. Therefore, since $I_\mathrm{e}$
 is isotropic,
we should substitute $\mu\frac{\partial I_\mathrm{e}}{\partial r}$
for $\D I_\mathrm{e}$ in Eq. (\ref{AI1del}).
Given Eq. (\ref{Blinear}) for $\B$, Eq. (\ref{AI1del})
can now be written as
  \begin{eqnarray}
   \delta I_1\, =\, -\mu\frac{1}{\lambda}
   \frac{\partial I_\mathrm{e}}{\partial r} +
   \frac{1}{\lambda}\int\!\!\!\int\!\!
   \frac{I_\mathrm{e}}{I_\mathrm{e}^\prime}\,
   R\left(\omega,\omega^\prime,\eta\right)
   \delta I_1^\prime\, d\Omega^\prime\, d\omega^\prime ,\label{EqIntGlob}\\
   \lambda\equiv\widetilde{l}^{-1}+\frac{1}{1-f_\mathrm{e}}
   \int\!\!\!\int\!\!
   \left(1-f_\mathrm{e}^\prime\right)\,
   R\left(\omega,\omega^\prime,\eta\right)\,
   d\Omega^\prime\, d\omega^\prime .\nonumber
  \end{eqnarray}
We derived an inhomogeneous Fredholm integral
equation of the second kind for the unknown function
$\delta I_1$. The free term of this equation is proportional to
$\mu$. Therefore, the solution should be proportional to $\mu$
and can be represented by relation (\ref{DImu}). Substituting
$\delta I =\mu G_1(\omega)$ and
$\delta I^\prime =\mu^\prime G_1(\omega^\prime)$
into Eq. (\ref{EqIntGlob}) yields
   \begin{equation}\label{EqIntG1}
    \mu G_1(\omega)\, =\, -\mu\frac{1}{\lambda}
    \frac{\partial I_\mathrm{e}}{\partial r} +
   \frac{1}{\lambda}\int\!\!\!\int\!\!
   \frac{I_\mathrm{e}}{I_\mathrm{e}^\prime}\,
   R\left(\omega,\omega^\prime,\eta\right)
   \mu^\prime\, G_1(\omega^\prime)\, d\Omega^\prime\, d\omega^\prime .
   \end{equation}
Here, when integrating over the angle
$(d\Omega^\prime =d\mu^\prime d\phi^\prime)$,
the variables $\mu^\prime$ and $\phi^\prime$ are closely intertwined
(see Eq. (\ref{etamu}) for $\eta$),
which complicates further simplifications
in the formal approach. However, we can use the
fact that the integral over the entire solid angle should
not depend on the choice of a coordinate system (for
more detail, see \cite{DAFK1959}.
So far we have measured the angles in the coordinate system
where the vertical axis is directed along the radius
vector and the azimuthal angle $\phi$ is counted off in the
plane perpendicular to it. Let us now direct the vertical
axis along the neutrino momentum vector before
the scattering and measure the azimuthal angle of the
scattered neutrino $\varphi$ in the plane perpendicular to it.
Then, $d\Omega^\prime =d\eta\, d\varphi^\prime$.
In addition, it is easy to obtain
the following expression for $\mu^\prime$:
   \begin{equation}\label{muprime}
    \mu^\prime =\mu\eta +
    \sqrt{1-\mu^{2}}\sqrt{1-\eta^{2}}\cos(\chi -\varphi^\prime)\, ,
   \end{equation}
where $\chi$ is the azimuthal angle of the radius vector
in the newly chosen frame. Let us now substitute
$\mu^\prime$ from Eq. (\ref{muprime}) into Eq. (\ref{EqIntG1}).
Since the integration of the
second term from Eq. (\ref{muprime}) over $\varphi^\prime$
in the range from 0 to $2\pi$ gives zero,
we obtain an integral equation for the
function $G_1(\omega)$ in final form
  \begin{eqnarray}
    & & G_1(\omega)\, =\, -\frac{1}{\lambda}
   \frac{\partial I_\mathrm{e}}{\partial r} +
   \frac{2\pi}{\lambda}\!\int_0^\infty\!
   \frac{I_\mathrm{e}}{I_\mathrm{e}^\prime}\,
   \int_{-1}^1\!\!\!\eta R\left(\omega,\omega^{\prime},\eta\right)d\eta
   G_1(\omega^\prime) d\omega^{\prime},\label{IntG1omeg}\\
   & & \lambda(\omega)\equiv\widetilde{l}^{-1}+\frac{2\pi}{1{-}f_\mathrm{e}}
   \int_0^\infty\!\!\left(1{-}f_\mathrm{e}^\prime\right)\,\!\!
   \int_{-1}^1\! R\left(\omega,\omega^\prime,\eta\right) d\eta
   \, d\omega^\prime . \label{IntG1lamb}
  \end{eqnarray}

If all elementary scattering processes are coherent,
then Eq. (\ref{IntG1omeg}) leads to a simple formula for $G_1(\omega)$
that, given Eq. (\ref{Rdelta}), can be written as
   \begin{eqnarray}
    & & G_1(\omega)\, =\, -\frac{1}
   {\widetilde{l}^{-1}+l_\mathrm{cs}^{-1}\, (1-\avr{\eta})}
   \frac{\partial I_\mathrm{e}}{\partial r}\, ,\label{lcogsct}\\
    & & \avr{\eta}\, =\, 2\pi l_\mathrm{cs}
   \!\int_{-1}^1\!\eta R_\mathrm{cs}(\omega,\eta)\, d\eta\, ,\quad
   l_\mathrm{cs}^{-1}\, =\, 2\pi\!\int_{-1}^1\! R_\mathrm{cs}
   (\omega,\eta)\, d\eta\, ,
   \end{eqnarray}
where $l_\mathrm{cs}$ is the neutrino mean free path with respect
to coherent scattering and \avr{\eta} is the cosine of the
scattering angle averaged over the scattering indicatrix.

The scattering cross section is
$\sigma_\mathrm{cs}=(l_\mathrm{cs}n_\mathrm{cs})^{-1}$,
where $n_\mathrm{cs}$ is the number of coherently scattering
particles per unit volume. Therefore, we can write
  $l_\mathrm{cs}^{-1}\, (1-\avr{\eta})=\sigma_\mathrm{cs}^\mathrm{t}
  n_\mathrm{cs}$.
The quantity
$\sigma_\mathrm{cs}^\mathrm{t}=\sigma_\mathrm{cs}\, (1-\avr{\eta})$
is called a transport cross section. In contrast to
Thomson photon scattering, the transport cross section
for coherent neutrino scattering depends significantly
on the neutrino energy $\omega$.
Let us now consider the more general case where
the coherent and incoherent neutrino scattering processes
are taken into account simultaneously. Let us
separate out the incoherent and coherent parts in the
scattering kernel $R$:
  \begin{equation}\label{RCognCog}
  R(\omega,\omega^\prime,\eta)\, =\,
  R_\mathrm{nc}(\omega,\omega^\prime,\eta)+
  R_\mathrm{cs}(\omega,\eta)\,\delta(\omega -\omega^\prime)\, .
  \end{equation}
After the substitution of this expression into Eq. (\ref{IntG1omeg})
and Eq. (\ref{IntG1lamb})
and elementary transformations, we obtain
  \begin{eqnarray}
   G_1(\omega)\, =\, -\frac{1}{\lambda_\mathrm{m}}
   \frac{\partial I_\mathrm{e}}{\partial r} +
   \frac{1}{\lambda_\mathrm{m}}\!\int_0^\infty\!
   \frac{I_\mathrm{e}}{I_\mathrm{e}^\prime}\,
   \Phi_1(\omega,\omega^\prime)\,
   G_1(\omega^\prime)\, d\omega^\prime ,\label{MIntG1omeg}\\
   \lambda_\mathrm{m}(\omega)\equiv\widetilde{l}^{-1}+
   l_\mathrm{cs}^{-1}\, (1-\avr{\eta}) +\frac{1}{1-f_\mathrm{e}}
   \int_0^\infty\!\!\!\left(1-f_\mathrm{e}^\prime\right)\,\!\!
   \Phi_0(\omega,\omega^\prime)
   \, d\omega^\prime , \label{MIntG1lamb}
    \end{eqnarray}
where we use the new notation
   \begin{equation}\label{Phi01}
   \Phi_0(\omega,\omega^\prime)=2\pi\!\!\int_{-1}^1\!
   R_\mathrm{nc}(\omega,\omega^\prime,\eta)d\eta\, ,\;\;
   \Phi_1(\omega,\omega^\prime)=2\pi\!\!\int_{-1}^1\!\eta R_\mathrm{nc}
   (\omega,\omega^\prime,\eta)d\eta\, .
   \end{equation}
We see that the sought-for integral equation did not
change in form. The situation was reduced only to
the replacement of $\lambda$ by the modified quantity $\lambda_\mathrm{m}$.
In the case of purely coherent scattering $(R_\mathrm{nc}=0)$,
we obtain solution given by Eq. (\ref{lcogsct}).
Obviously, the integral equation for antineutrinos
has exactly the same form; we only need to change
the sign of the chemical potential $\psi$ in $I_\mathrm{e}$ and use the
corresponding expression for the scattering kernel $R$
and mean free paths $l$.

\subsubsection*{
The leptonic charge diffusion equation
}

The leptonic charge $\Lambda$ is equal to the difference
between the numbers of leptons (electrons plus
neutrinos: $n_{-} + n_\nu$) and antileptons (positrons plus
antineutrinos: $n_{+} + n_{\overline{\nu}}$) in the system
and is conserved
in any elementary interactions of neutrinos and
antineutrinos with matter. For the specific leptonic
charge, we have
  \begin{equation}\label{Lepchar}
  \Lambda\, =\,\frac{m_\mathrm{u}}{\rho}\left(n_{-} - n_{+}
  \, +\, n_\nu - n_{\overline{\nu}}\right)\, ,
  \end{equation}
where $m_{\mathrm{u}}$ is the atomic mass unit. The difference between
the numbers of electrons and positrons should
satisfy the electrical neutrality condition
  \begin{equation}\label{electrn}
  n_{-} - n_{+}\, =\, n_{\mathrm{p}}\, ,
  \end{equation}
where $n_{\mathrm{p}}$ is the total number of protons (both free
ones and those bound in atomic nuclei), which together
with the number of neutrons defines the matter
density $\rho = m_{\mathrm{u}}(n_{\mathrm{p}}+n_{\mathrm{n}})$.
The specific numbers of
protons $(m_\mathrm{u}n_{\mathrm{p}}/\rho)$
and neutrons $(m_\mathrm{u}n_{\mathrm{n}}/\rho)$ per nucleon
can change only through reactions involving
neutrinos and antineutrinos.

Let us consider the leptonic charge transport
equation in the comoving frame (we omit the subscript
0 as before). Clearly, the equilibrium leptonic
charge $\Lambda_{\mathrm{e}}$ can change only through a deviation
of the difference $n_\nu - n_{\overline{\nu}}\,$
from its equilibrium value.
Therefore,
the derivative $d\Lambda_{\mathrm{e}}/dt$ should be equal to the rate
of change in
  $(n_\nu -n_{\nu\mathrm{e}}) -
  (n_{\overline{\nu}}-n_{\overline{\nu}\mathrm{e}})$.
However, the
latter is defined by the right-hand sides of the neutrino
and antineutrino number transport equations, respectively.
These equations can be obtained from the
neutrino energy transfer equation (\ref{Transeq}) by substituting
\INU\ and \INUA\ (and their equilibrium values)
for $\INU /\omega_\nu$ and $\INUA /\omega_{\overline{\nu}}\,$,
respectively. The transformation of the righthand
sides of the neutrino and antineutrino number
transport equations is identical to that made above for
the energy transfer equation. As a result, we obtain
   \begin{equation}\label{Diffuselc}
   \frac{d\Lambda_{\mathrm{e}}}{dt}\, =\, -
   \frac{4\pi}{3}\frac{\nucmu}{\rho}
   \frac{1}{r^{2}}\frac{\partial}{\partial r}\left[r^{2}
   \left(\int_{0}^{\infty }\!\! G_{1\nu}\,\frac{d\omega_\nu}{\omega_\nu}\, -
   \int_{0}^{\infty }\!\! G_{1\overline{\nu}}\,
   \frac{d\omega_{\overline{\nu}}}{\omega_{\overline{\nu}}}\,
   \right)\right]\, .
   \end{equation}
where $G_{1\nu}\,$ and $G_{1\overline{\nu}}\,$
 are the solutions to the integral
equation (\ref{MIntG1omeg}) for neutrinos and antineutrinos, respectively.
Thus, describing the leptonic charge diffusion
does not require solving a new integral equation.

\subsubsection*{Transformation of the integral equation}

If the derivative $\partial I_\mathrm{e}/\partial r$
as a function of the neutrino
energy is known, then the functions $G_{1\nu}(\omega)\,$  and
$G_{1\overline{\nu}}(\omega)\,$, along with the energy and
leptonic charge fluxes, can be determined by numerically solving
Eq. (\ref{MIntG1omeg}). Although this straightforward approach
is possible, for the reliability of the numerical calculations
and for the physical understanding, it is
appropriate to separate out the contributions from the
temperature and chemical potential gradients to these
fluxes.

By introducing a new sought-for function $F_1(\omega)\,$,
\begin{equation}\label{GtoF}
  G_1(\omega)\,  =\, -\frac{\omega^3}{c^2 h^3}\,
  \frac{1}{\lambda_\mathrm{m}(\omega)}\, F_1(\omega)
\end{equation}
 we transform Eq.(\ref{MIntG1omeg}) to
 \begin{equation}\label{IntF}
   F_1(\omega) =\frac{\partial f_\mathrm{e}}
   {\partial r} + f_\mathrm{e}
   \!\int_0^\infty\! \Phi_1(\omega,\omega^\prime)F_1(\omega^\prime)
   \frac{d\omega^\prime}{f_\mathrm{e}^\prime\,
   \lambda_\mathrm{m}(\omega^\prime)}\, ,
 \end{equation}
The derivative of the equilibrium filling factor $f_\mathrm{e}$
can be represented as
\begin{equation}
\frac{\partial\feq}{\partial
r}=\feq(1-\feq)\left(\frac{\omega}{kT^2}\frac{\partial T}{\partial r}
+\frac{\partial\psi}{\partial r}\right).
\label{dfedr}
\end{equation}
Therefore, we will seek a solution to Eq. (\ref{IntF}) in the form
\begin{equation}
 F_1(\omega)=\feq\left[\left(\frac{1}{kT^{2}}\frac{\partial T}
 {\partial r}\right) \gT(\omega)+\left(\frac{\partial\psi}
 {\partial r}\right)\gpsi(\omega)\right]. \label{F1form}
\end{equation}
We substitute this expression into Eq. (\ref{IntF}), collect the
terms containing
$\partial T/\partial r$ and $\partial\psi/\partial r$, and set them
equal to zero separately. As a result, we obtain the following
equations for the functions $\gT(\omega)$ and $\gpsi(\omega)$:
\begin{equation}
\left\{
\begin{aligned}
\label{gTgpsi}
\gT(\omega)&=\omega(1{-}\feq)+\int_0^\infty\!\!
\Phi_1(\omega,\omega^\prime)\,\gT(\omega^\prime)
\frac{d\omega^\prime}{\lamb(\omega^\prime)}\, ,\\
\gpsi(\omega)&=(1{-}\feq)+\int_0^\infty\!\!
\Phi_1(\omega,\omega^\prime)\,\gpsi(\omega^\prime)
\frac{d\omega^\prime}{\lamb(\omega^\prime)}\, .
\end{aligned}
\right.
\end{equation}
Thus, we have a pair of separated integral equations
with identical kernels, but with different free terms.
Generally, these equations have to be solved numerically
for neutrinos and antineutrinos at given temperature
and chemical potential. The method of numerical
solution that is used in our current calculations of
gravitational collapse is described in our paper
\cite{YudNad2008}, see Appendix B there.

\subsubsection*{
The neutrino heat conduction equations
}

Let us write the system of NHC equations that was first derived
$\sim 40$ years years ago \cite{ImshNad1972}.
%
\begin{align}
\label{NHCE1}
 & \frac{d r}{d t}=u,\quad
   \frac{1}{\rho}=\frac{4\pi}{3}\frac{\partial r^{3}}{\partial m}\, ,\\
 & \frac{d u}{d t}=-4\pi r^{2}\frac{\partial}{\partial m}
   \Bigl(P+P_{\nu}\Bigr)-\frac{G m}{r^{2}}\, ,\\
 & \frac{d}{dt}\left(E+\frac{U_{\nu}}{\rho}\right)+
   \Bigl(P+P_{\nu}\Bigr)\frac{d}{dt}\left(\frac{1}{\rho}\right)
   =-4\pi\frac{\partial}{\partial m}\Bigl(r^{2} H_{\nu}\Bigr)\,
   ,\\ \label{NHCE4}
 & \frac{d\Lambda_\nu}{dt}+4\pi\,\nucmu\frac{\partial}
   {\partial m}\Bigl(r^{2}F_{\nu}\Bigr)=0\, ,
 \end{align}
where $m$ is the mass coordinate, while $P$ and $E$ are
the matter pressure and energy; $P_\nu$ and $U_\nu$ are the
sums of the equilibrium neutrino and antineutrino
pressures and energy densities defined by Eqs. (\ref{UK0E})
with $\psi$ replaced by $-\psi$ for antineutrinos. Here, we use
a new notation for the neutrino pressure, $P_\nu\equiv\KNUOE\,$.
The equilibrium specific leptonic charge $\Lambda_\nu$ can be
written as
 \begin{equation}\label{Lambdsp}
 \Lambda_\nu\, =\, (Y_- - Y_+) + (Y_\nu - Y_{\overline{\nu}})\, ,
 \end{equation}
where $Y_-$, $Y_+$, $Y_\nu$, and $Y_{\overline{\nu}}$ are
the equilibrium numbers
of electrons, positrons, neutrinos, and antineutrinos
per nucleon $(Y =n\nucmu/\rho)$. We should use Eq. (\ref{NNUE})
for $Y_\nu$ and the same formula for $Y_{\overline{\nu}}$
with $\psi$ replaced
by $-\psi$.
From the electrical neutrality condition, we
have $Y_- - Y_+ = Y_\mathrm{p}=1-Y_\mathrm{n}$,
where $Y_\mathrm{p}$ and $Y_\mathrm{n}$ are the
total specific numbers of protons and neutrons in the
system.

The neutrino energy, $H_{\nu}$, and leptonic change, $F_{\nu}$ ,
fluxes can be written as
\begin{eqnarray}
 H_{\nu}=\frac{4\pi}{3}\!\int_0^\infty\!(G_{1\nu}+G_{1\overline{\nu}})d\omega
 = -\frac{4\pi}{3h^{3}c^{2}}
 \!\int_0^\infty\!\!\left(\frac{F_{1\nu}}{\lambn}
 + \frac{F_{1\overline{\nu}}}{\lamban}\right)\!\omega^{3}d\omega\, ,\\
 F_{\nu}=\frac{4\pi}{3}\!\int_0^\infty\!(G_{1\nu}-G_{1\overline{\nu}})
 \frac{d\omega}{\omega} =
 -\frac{4\pi}{3h^{3}c^{2}}\!
 \int_0^\infty\!\!\left(\frac{F_{1\nu}}{\lambn}
 - \frac{F_{1\overline{\nu}}}{\lamban}\right)\!\omega^{2}d\omega\, .
\end{eqnarray}
Substituting here $F_1$ from Eq. (\ref{F1form}), we express the
fluxes in terms of the temperature and chemical potential
gradients:
\begin{eqnarray}
  H_{\nu} = -\frac{4\pi}{3h^{3}c^{2}}
  \left[(A_\nu + A_{\overline{\nu}})\frac{1}{kT^2}\frac{\partial T}{\partial r}
  + (B_\nu - B_{\overline{\nu}})\frac{\partial\psi_\nu}{\partial r}\right],
  \label{Hgrad}\\
  F_{\nu} = -\frac{4\pi}{3h^{3}c^{2}}
    \left[(C_\nu - C_{\overline{\nu}})\frac{1}{kT^2}\frac{\partial T}{\partial r}
  + (D_\nu + D_{\overline{\nu}})\frac{\partial\psi_\nu}{\partial r}\right] .
   \label{Fgrad}
  \end{eqnarray}
where the derivative $\partial\psi_{\overline\nu} /\partial r$
was replaced by $-\partial\psi_\nu /\partial r$.
The kinetic coefficients $A$, $B$, $C$ and $D$
are specified by the integrals
\begin{equation}\label{ABCD}
\begin{aligned}
 A &=\!\!\int_0^\infty\!\! \frac{f_{\mathrm{e}}}{\lamb}\gT\,\omega^3d\omega
    \, ,\qquad
 B &= \!\!\int_0^\infty\!\! \frac{f_{\mathrm{e}}}{\lamb}\gpsi
    \,\omega^3d\omega\, , \\
 C &=\!\!\int_0^\infty\!\! \frac{f_{\mathrm{e}}}{\lamb}\gT\,\omega^2d\omega
    \, ,\qquad
 D &= \!\!\int_0^\infty\!\! \frac{f_{\mathrm{e}}}{\lamb}\gpsi
    \,\omega^2d\omega\, ,
\end{aligned}
\end{equation}
where the absence of subscripts $\nu$ and $\overline{\nu}$ indicates
that the expressions are applicable to both neutrinos and
antineutrinos.

Owing to the identity $B\equiv C$, we have
 $(B_\nu - B_{\overline{\nu}})=(C_\nu - C_{\overline{\nu}})$.
This identity is not accidental, but is an expression of the Onsager
symmetry principle for the kinetic coefficients. Its
validity for incoherent neutrino scattering is ensured
by the symmetry property for the scattering kernel
expressed by Eq. (\ref{Rexch}). In the case of purely coherent
scattering and in the absence of scattering, we have
$R_\mathrm{nc}(\omega,\omega^\prime,\eta)=0$,
$\Phi_1(\omega,\omega^\prime)=0$
and the satisfaction of this identity is trivial,
since $\gT =\omega\, (1-f_\mathrm{e})$, $\gpsi =(1-f_\mathrm{e})$.
The validity of the identity for incoherent
scattering is proved in paper \cite{YudNad2008} (see Appendix A there).

When deriving the system of the NHC
equations $(\ref{NHCE1}{-}\ref{NHCE4})$,
we used a number of assumptions
regarding the neutrino radiation and matter
properties: the smallness of the neutrino mean free
path compared to the characteristic size of the problem,
the smallness of the velocities in matter compared
to the speed of light, etc. The conditions that
the quantities appearing in the theory should satisfy
follow from these assumptions. Thus, for example,
since the correction $\delta I_1$ to the equilibrium intensity
$\IE$ in Eq. $(\ref{I1PDEL})$ should be small, we obtain
constraints on the leptonic number and energy fluxes
which formally can be written as
\begin{equation}
\label{FirstCondition} \left|\frac{F_{\nu}}{c
n_{\nu}}\right|\ll 1,\quad \left|\frac{H_{\nu}}{c
U_{\nu}}\right|\ll 1.
\end{equation}
In addition, we can obtain conditions for the rates
of change in leptonic change $\Lambda_\nu$ and total specific
energy $E_{\mathrm{tot}}=E+\frac{U_{\nu}}{\rho}$:
\begin{equation}
\label{SecondCondition} \frac{\avr{l_\nu}\,}{c
\Lambda_\nu}\left|\frac{d \Lambda_\nu}{d t}\right|\ll 1,\quad
\frac{\avr{l_\nu}}{c E_{\mathrm{tot}}}\left|\frac{d
E_{\mathrm{tot}}}{d t}\right|\ll 1,
\end{equation}
where \avr{l_\nu} is the mean free path.
Conditions $(\ref{SecondCondition})$
have a simple meaning: they require that the relative
change in the quantities under consideration be small
in the characteristic time of a free neutrino transit
$\tau =\avr{l_\nu}/c$.
These inequalities exhaust the applicability
conditions for the NHC approximation.

When using the NHC for calculating the neutrino transport
one has to define domain of its applicability by specifying
the values $(<1)$ for the left hand sides of the inequalities
(\ref{FirstCondition}) and (\ref{SecondCondition}).
The outer boundary of the domain, determined this way in calculations,
is not necessarily coincides with the neutrinosphere
analogous to photosphere for common stars.

\subsubsection*{Concluding remarks}
Remarkably, in the NHC approximation,
the scattering effects enter {\em only\/} via
the zeroth and first moments
$\Phi_{0,1}(\omega,\omega^\prime)$ of the expansion
of the kernel $R(\omega,\omega^\prime,\eta)$ in terms
of Legendre polynomials.
It was pointed out previously
(see, e.g., papers \cite{MezzBruenn1993,SmitCern1996})
that the results of numerical hydrodynamic calculations of neutrino
transport are insensitive to whether the next terms in the
expansion of the scattering kernel are included.
Such insensitivity indicates that the physical conditions in a
collapsing stellar core are close to those needed for the
NHC  regime   to be established.

Here we disregarded the processes in which both neutrinos and
antineutrinos are involved, such as the annihilation of
electron–positron pairs, the neutrino decay of plasmons,
the bremsstrahlung of neutrino
pairs, etc. Although the role of these processes is minor
during the gravitational collapse of stellar cores,
their description in the NHC theory
is definitely of interest. The main modification
here is related to the fact that the transport equations
(\ref{Transeq}) for neutrinos and antineutrinos lose their
independence and form a system of two equations
with the neutrino and antineutrino intensities appearing
simultaneously on their right-hand sides. As a
result, when calculating the kinetic coefficients, we
would have to solve not two pairs of separated integral equations
(\ref{gTgpsi}), but instead two systems of two coupled equations
in each: one for $g_{T\nu}$, $g_{T\overline{\nu}}$
and the other for $g_{\psi\nu}$, $g_{\psi\overline{\nu}}$.

The NHC theory consistently describes fluxes of energy and
lepton charge emerging from the neutrino opaque core.
The fluxes are proportional to the
gradients of temperature and neutrino chemical
potential.

Incoherent neutrino scattering enters the NHC equations
through 0-th and 1-st moments of
the Legendre expansion of scattering kernel.
Coherent scattering is described by the transport
cross-section algorithm.

\section*{Realization of the NHC theory in calculations
 of gravitational collapses}

\subsection*{Rotating stars}
 Let us consider an axially symmetric rotating star.
 The 2D-NHC equations for this case were derived
 in PhD Thesis \cite{Yud2009}.

 The equations of motion and continuity read
\begin{equation}
\left\{
\begin{aligned}
& \frac{d V_{r}}{d t}-V_{\varphi}\,\omega =-\frac{1}{\rho}\frac{\partial}{\partial r}
\bigl(P+P_{\nu}\bigr)-\frac{\partial\Phi}{\partial r}\, ,\\
& \frac{d V_{\varphi}}{d t}+V_{r}\,\omega=0\, ,\\
& \frac{d V_{z}}{d t}=-\frac{1}{\rho}\frac{\partial}{\partial
z}\bigl(P+P_{\nu}\bigr)-\frac{\partial\Phi}{\partial z}\, ,
\end{aligned}
\right.\label{2D-NHC-Motion}
\end{equation}
\begin{equation}
\frac{\partial\rho}{\partial
t}+\frac{1}{r}\frac{\partial}{\partial r}\bigl(r\rho
V_{r}\bigr)+\frac{\partial}{\partial z}\bigl(\rho
V_{z}\bigr)=0\, ,\label{2D-NHC-Continuity}
\end{equation}
where $r$ is cylindrical radius, and the Lagrangian operator
$$\frac{d}{d t}=\frac{\partial}{\partial t}
+V_{r}\frac{\partial}{\partial r}+V_{z}\frac{\partial}{\partial z}\, .$$
In this section  $\omega =\omega(r,z)$ is the angular
velocity of rotation and $\Phi$ is the gravitational potential
that is described by Poisson’s equation
\begin{equation}\label{Poisson}
    \Delta\Phi\equiv \frac{\partial}{\partial r}
    \left(r\frac{\partial \Phi}{\partial r}\right)
    +\frac{\partial^2\Phi}{\partial z^2}=-4\pi G\rho\, .
\end{equation}

The equations of energy and of leptonic charge conservation are given by
\begin{equation}
\frac{d}{d
t}\left(E{+}\frac{U_{\nu}}{\rho}\right)+\bigl(P{+}P_{\nu}\bigr)\frac{d}{d
t}\left(\frac{1}{\rho}\right)=-\frac{1}{\rho}\left[\frac{1}{r}\frac{\partial}{\partial
r}\bigl(r H_{\nu r}\bigr)+\frac{\partial H_{\nu z}}{\partial
z}\right].\label{2D-NHC-Energy}
\end{equation}

\begin{equation}
\frac{d\Lambda_\nu}{d t}=-\frac{\WID{m}{u}}{\rho}
\left[\frac{1}{r}\frac{\partial}{\partial r}\bigl(r F_{\nu
r}\bigr)+\frac{\partial F_{\nu z}}{\partial
z}\right].\label{2D-NHC-Lepton}
\end{equation}
The leptonic charge $\Lambda_\nu$ is given by Eq. (\ref{Lepchar}).
In Eqs. (\ref{2D-NHC-Motion}) and (\ref{2D-NHC-Energy})
$P_{\nu}$ and $U_{\nu}$ are total equilibrium pressure and energy density
of neutrino and antineutrino. The fluxes of energy $H_{\nu r}$,
$H_{\nu z}$ and of leptonic charge $F_{\nu r}$, $F_{\nu z}$
in Eqs. (\ref{2D-NHC-Energy}) and (\ref{2D-NHC-Lepton})
are given by
\begin{eqnarray}
  H_{\nu r} = -\frac{4\pi}{3h^{3}c^{2}}
  \left[(A_\nu + A_{\overline{\nu}})\frac{1}{kT^2}\frac{\partial T}{\partial r}
  + (B_\nu - B_{\overline{\nu}})\frac{\partial\psi_\nu}{\partial r}\right],
  \label{Hgradr}\\
  H_{\nu z} = -\frac{4\pi}{3h^{3}c^{2}}
  \left[(A_\nu + A_{\overline{\nu}})\frac{1}{kT^2}\frac{\partial T}{\partial z}
  + (B_\nu - B_{\overline{\nu}})\frac{\partial\psi_\nu}{\partial z}\right],
  \label{Hgradz}
  \end{eqnarray}
\begin{eqnarray}
 F_{\nu r} = -\frac{4\pi}{3h^{3}c^{2}}
    \left[(C_\nu - C_{\overline{\nu}})\frac{1}{kT^2}\frac{\partial T}{\partial r}
  + (D_\nu + D_{\overline{\nu}})\frac{\partial\psi_\nu}{\partial r}\right] ,
   \label{Fgradr}\\
  F_{\nu z} = -\frac{4\pi}{3h^{3}c^{2}}
    \left[(C_\nu - C_{\overline{\nu}})\frac{1}{kT^2}\frac{\partial T}{\partial z}
  + (D_\nu + D_{\overline{\nu}})\frac{\partial\psi_\nu}{\partial z}\right] .
   \label{Fgradz}
  \end{eqnarray}

 We see that in case of 2D symmetry the kinetic coefficients
 $A,\, D$ and $B\equiv C$ are determined
 exactly as in the case of spherical symmetry given by Eqs. (\ref{ABCD}).
 It is clear that this should be true also in the 3D case.

\subsection*{Examples of the NHC calculations}
\begin{figure}[htb!]
\centerline{\includegraphics[clip,width=0.8\textwidth]{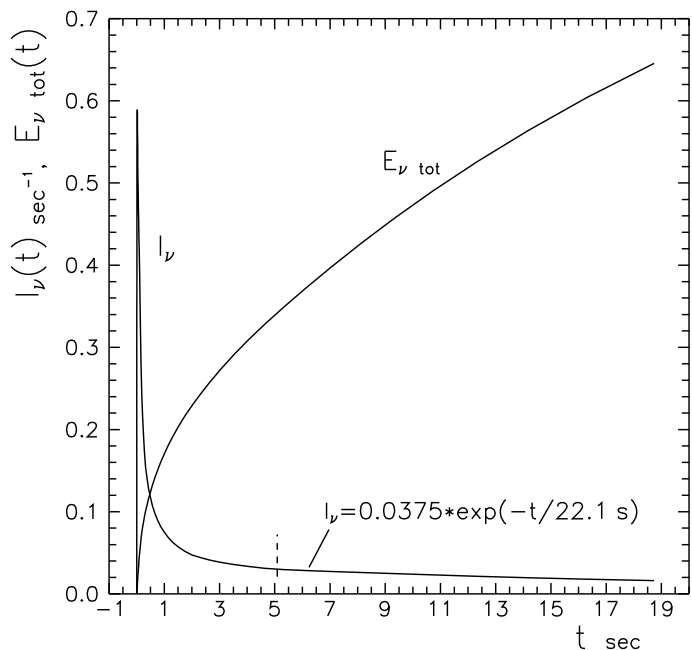}}
\caption{The normalized neutrino light curve
$l_\nu(t)=L_{\nu\mathrm{tot}}/{\cal E}_{\nu\mathrm{tot}}$
 and integrated energy of the neutrino flux
 $E_{\nu\mathrm{tot}}(t)=\int_0^t\!l_\nu\, dt$,
 $E_{\nu\mathrm{tot}}(\infty)=1$.
 ${\cal E}_{\nu\mathrm{tot}}$ is the total energy emitted
  by neutrinos of all the flavors.
 Adapted from paper \cite{Nad78}.
}
\label{LnuNorm}
\end{figure}

For the first time the NHC approach to gravitational collapse
was used in 1978 \cite{Nad78}. Figure \ref{LnuNorm} shows
the resulting neutrino light curve $L_{\nu\mathrm{tot}}(t)$
of all the neutrino and antineutrino flavors.


 The calculations were performed for a \NMS{1.82}
 iron core surrounded with a \NMS{0.18} oxygen envelope.

 Full energy carried away by neutrinos
 ${\cal E}_{\nu\mathrm{tot}}=\int^\infty_0 L_{\nu\mathrm{tot}}(t)\mathrm{dt}$
 turned out to be \xmn{5.3}{53}erg.
 During the first $\sim 100\,$ms of the collapse,
 $L_{\nu\mathrm{tot}}(t)$ is building up mostly by the electron neutrino
 from neutronization of stellar matter.
 However, later on
 when the neutrino flux comes from neutrinosphere,
 an approximate equidistribution over the neutrino flavors set in.

 The electron antineutrino spectrum was estimated to be
 the Fermi--Dirac distribution with zero chemical potential
 and temperature $\sim 4\,$MeV,
 the corresponding  mean energy of emitted electron antineutrino
 \avr{E_{\bar\nu\mathrm e}} being $\sim 12\,$MeV.
 Such individual $\bar\nu_{\mathrm e}$ energy and predicted
 long time (10--20)$\,$s of the neutrino light curve decay
 were confirmed by the underground neutrino detectors which
 observed the the neutrino signal from supernova 1987A
 in the Large Magellanic Cloud \cite{ImNad89Sn87A,NadIm2005}.

 Detailed study of spherically symmetrical gravitational collapse
 was undertaken in PhD Thesis \cite{Yud2009}.
 The NHC was used in central neutrino-opaque region of contracting
 stellar core. The radius of outer boundary of the region
 was repeatedly recalculated in accordance with inequalities
 (\ref{FirstCondition}, \ref{SecondCondition}) to ensure a smooth
 coupling with outermost semi-transparent for neutrino envelope.
 There the neutrino transfer equation (\ref{Transeq}) was calculated by using
 a special finite difference  scheme (in space and angle coordinates)
 that in the limit of large neutrino ``optical'' depth gives
 the neutrino fluxes exactly the same as predicted by the NHC.

\begin{figure}[htb!]
\centerline{\includegraphics[clip,width=0.8\textwidth]{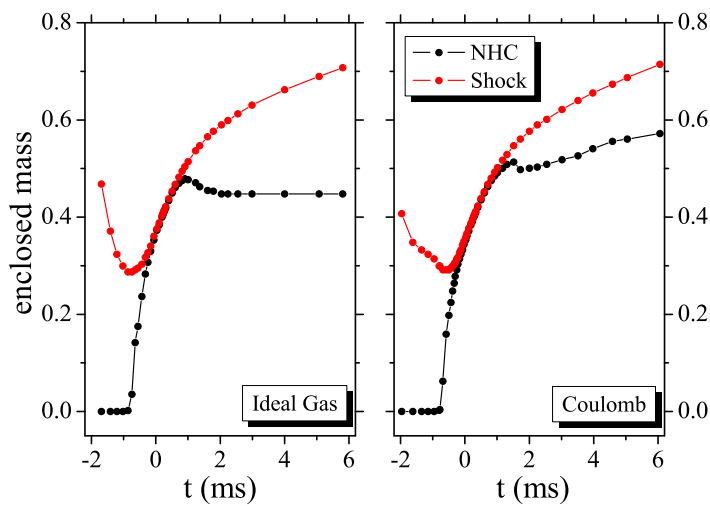}}
\caption{Gravitational collapse of a \NMS{2} iron core: the dimensionless
masses ($m/\NMS{2}$) enclosed within the position of maximum velocity
of infalling matter (red dots)
and of outer NHC boundary (black dots) versus time measured from the moment
of bounce $(t=0)$.
}
\label{EncMass}
\end{figure}
Figure \ref{EncMass} shows how the mass coordinate of maximum velocity
of infalling matter and that of the position of outer boundary of
the NHC domain vary with time during the first milliseconds before
and after bounce. At the bounce the shock wave first appears and the
maximum velocity of infalling matter coincides with the outer edge
of the shock front.
The NHC first appears at about -1$\,$ms before bounce.
However, in 2$\,$ms at ($t\approx 1\,$ms) it comprises already about
a half of the total mass and controls major part of total neutrino
energy and leptonic charge fluxes.

The neutrino transport above the NHC black-dot-line is calculated
by using  aforementioned difference scheme.
The left and right panels of Fig. \ref{EncMass} depict two versions
of calculations which have some differences in equation of state
(ideal gas or with coulomb interaction allowed for) and
in the rigidity adopted for inequalities (\ref{FirstCondition}) which
define the NHC outer boundary. One can observe that after bounce
($t>0$) the position of accreting shock wave is practically
unaffected by such details.

\begin{figure}[htb!]
\centerline{\includegraphics[clip,width=0.8\textwidth]{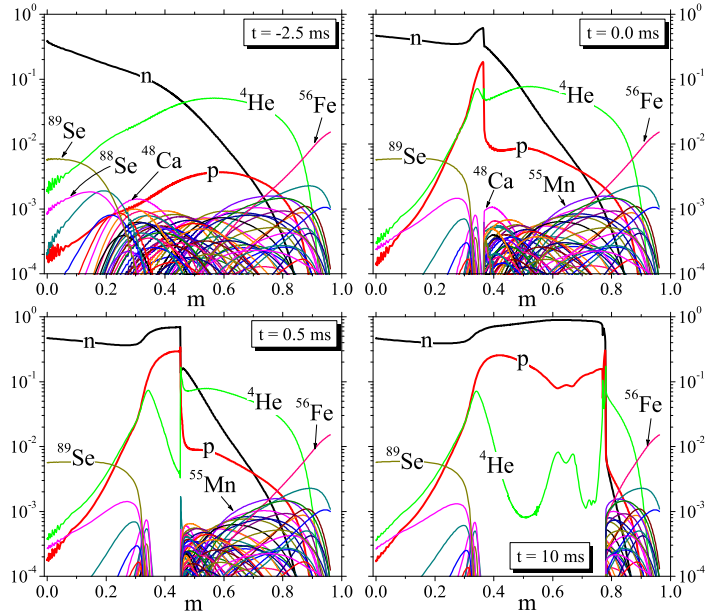}}
\caption{Composition under conditions of nuclear statistical equilibrium
versus dimensionless mass coordinate at different
times for calculations shown in Fig.$\,$\ref{EncMass}.
}
\label{Yieps}
\end{figure}

Figure \ref{Yieps} shows composition in terms of $Y_i=\nucmu n_i/\rho$
where $n_i$ is the number of nuclides per unit volume with $i=$ n, p,
\isn{He}{4},\ldots, \isn{Kr}{86} --- in total 137 stable and unstable
most representative isotopes of the iron peak elements \cite{NadYud04}.
At $t=-2.5\,$ms (top left panel), central core of a mass $\approx$\NMS{1}
 becomes strongly neutronized. The shock wave first appears
at bounce at m$\approx 0.35$  (\NMS{0.7}, top right panel) and begins
steadily to dissociate heavy nuclides mostly into
n, p, and \isn{He}{4} thereby creating a growing n-p-\isn{He}{4}
shell that in 10$\,$ms contains already about \NMS{0.9} (bottom panels).

\section*{Equation of state for nuclear statistical equilibrium}
At temperatures $T_9 \gtrsim 4$ the  thermonuclear
reactions turn out to be so fast that they can establish the state
of nuclear statistical equilibrium (NSE). The most effective in
launching such a process are the direct and inverse ($\gamma,$p),
($\gamma,$n), (n,p), and ($\gamma,\alpha$) reactions. As a result,
the NSE abundances $n_{A,Z}$ of all the nuclides $(A,Z)$
depend on only three external parameters: temperature $T$, density $\rho$
and a ratio $\theta$ of total number neutrons N$_{\mathrm n}$
(free and bounded in nuclei) to that of protons N$_{\mathrm p}$:
\begin{equation}\label{theta}
 \theta = \frac{{\mathrm N}_{\mathrm n}}{{\mathrm N}_{\mathrm p}}
 =\frac{n_{\mathrm n}+\sum_{Z\geqslant 1,A\geqslant 2}(A-Z)n_{A,Z}}
  {n_{\mathrm p}+\sum_{Z\geqslant 1,A\geqslant 2}Zn_{A,Z}}\, ,
\end{equation}
where $n_{\mathrm n}$ and $n_{\mathrm p}$ are the number densities
of free neutrons and protons, respectively.
Using the definition of density
\begin{equation}\label{density}
    \rho = \nucmu\, ({\mathrm N}_{\mathrm n}+{\mathrm N}_{\mathrm p})
\end{equation}
and taking into account equations of statistical equilibrium
connecting $n_{A,Z}$ with $n_{\mathrm n}$ and $n_{\mathrm p}$
(see paper \cite{NadYud04} and references therein) one can
calculate all thermodynamical quantities of nuclear component,
such as pressure $P_{\mathrm{nuc}}$,
specific energy $E_{\mathrm{nuc}}$, and entropy $S_{\mathrm{nuc}}$.
After addition of same quantities for the black body radiation
and electron-positron
components (see paper \cite{epeos96}) we obtain three parametric
equation of state (EOS) appropriate for calculating the gravitational collapse
of stellar cores:
\begin{equation}\label{PES}
 P=P(T,\rho,\theta),\quad  E=E(T,\rho,\theta),\quad S=S(T,\rho,\theta)\, .
\end{equation}

 In 1946 F.~Hoyle first showed \cite{Hoyle1946}
 that a decomposition of matter under
 the conditions of  NSE can lead to the collapse of the central
 stellar core.
The most important EOS property is the adiabatic index $\gamma$.
Its critical value 4/3 serves as the boundary between hydrostatically
stable stellar cores $(\avr{\gamma} > 4/3)$ and unstable ones
$(\avr{\gamma} < 4/3), \avr{\gamma}$
being properly averaged $\gamma$ over the hydrostatic structure of a core.
\begin{equation}\label{Gamma}
\gamma = \left(\frac{\partial \log P}{\partial \log\rho}\right)_{S,\theta}\, .
\end{equation}

\begin{figure}[htb!]
\centerline{\includegraphics[clip,width=0.8\textwidth]{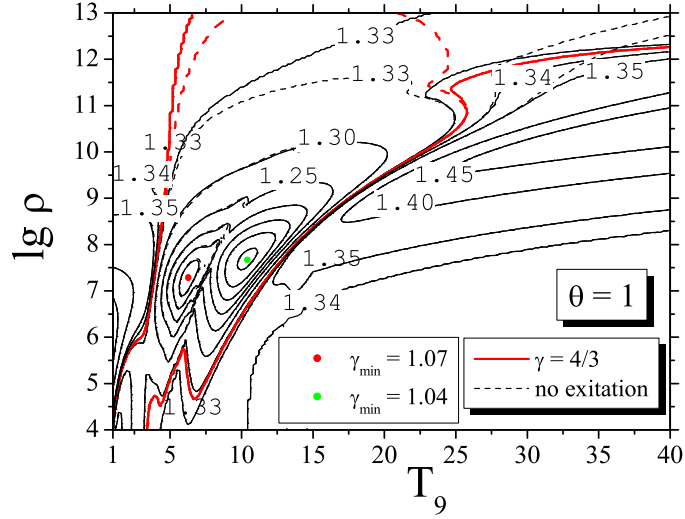}}
\caption{Lines of constant adiabatic index $\gamma$ for $\theta =1$
(Adapted from paper \cite{NadYud04}).
}
\label{Gam1eps}
\end{figure}

\begin{figure}[htb!]
\centerline{\includegraphics[clip,width=0.8\textwidth]{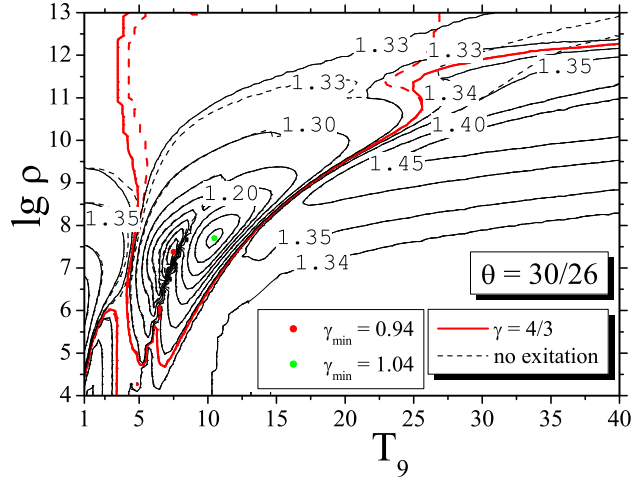}}
\caption{Same as in Fig.$\,$\ref{Gam1eps} but for $\theta =30/26$.
}
\label{GamFeeps}
\end{figure}

 Figures \ref{Gam1eps} and \ref{GamFeeps} show the lines of constant $\gamma$
 on the $T-\rho$ plane for $\theta=1$ and 30/26, respectively.
At low and high densities, $\gamma$ becomes less than 4/3
due to the appearance of electron–positron pairs and
the dissociation of nuclides into $\alpha$ particles and free
nucleons, respectively.

For $\theta=1$, the ravine of instability $\gamma < 4/3$
is a singly connected domain: one of the above effects
smoothly replaces the other. However, for $\theta=30/26$,
a narrow  neck that separates the instability
ravine into two parts arises at $3\lesssim T_9 \lesssim 4$.
Height of the neck is however small $(\gamma - 4/3 \approx 0.002)$.

This happens because the most abundant nuclides
at such temperatures are \isn{Ni}{56} for $\theta=1$ and \isn{Fe}{56}
for $\theta=30/26$; the binding energy of the latter nuclide is
higher by 8.4 MeV. As a result, the \isn{Fe}{56} dissociation
begins at a higher temperature than does the \isn{Ni}{56} one,
when electron–positron pairs contribute to lowering $\gamma$
not so strongly.
At $\theta$ that are appreciably larger than 30/26$\approx$1.154,
for instance such as 1.5,
the instability ravine again becomes singly connected domain.
The reason is the same: the most abundant
nuclides have lower binding energies
than that for \isn{Fe}{56}.

Each figure is a superposition of two sets of lines of constant $\gamma$
calculated with (solid lines) and without (dashed lines) nuclear
excited states as discussed in paper \cite{NadYud04}.
At low densities $(\rho\lesssim 10^9\,$g$\,$cm$^{-3}$),
both sets of lines virtually merge together, suggesting
that $\gamma$ is weakly sensitive to the nuclide excitation
parameters. However, at densities $\rho\gtrsim 10^{10}\,$g$\,$cm$^{-3}$,
the excitation of nuclides causes mainly a reduction
in $\gamma$, which is accompanied by an expansion of the
instability ravine.
The boundaries of this region are marked by the heavy solid
and dashed lines on which $\gamma$ is exactly equal to 4/3.

A characteristic feature of the function $\gamma(T,\rho,\theta)$
is the existence of two deep minima that are clearly seen
in Figs. \ref{Gam1eps} and \ref{GamFeeps}.
The left minimum results from the dissociation of
nuclides into $\alpha$ particles and free nucleons in the presence
of electron–positron pairs, while the right one arises from
the dissociation of $\alpha$ particles into free nucleons.
In case of \isn{Fe}{56} dominated NSE matter (Fig. \ref{GamFeeps})
the minima are $\gamma_\indis{min}=0.94$ (left) and
$\gamma_\indis{min}=1.04$ (right).

It is worth to mention that the NSE EOS obtained
in paper \cite{ImNad1965} resulted in the $\gamma_\indis{min}$ values
0.98 and 1.06, close to that shown in Fig. \ref{GamFeeps}.
It is the NSE EOS that was used in calculations of stellar core collapse
resulting in the neutrino light curve shown in Fig. \ref{LnuNorm}
above. These authors assumed the set of iron-group elements
to be represented by one isotope, \isn{Fe}{56}, with only 7 excited states.
In addition, a thermodynamic equilibrium of the $\beta$-processes
with a zero neutrino chemical potential was assumed,
which allowed to exclude $\theta$ as the third independent variable.
Another approach to the problem was implemented in paper
\cite{ImshChech1970} in which the kinetic equilibrium approximation
for the $\beta$-processes to determine $\theta$ as a function of the
temperature and density was used.

\begin{figure}[htb!]
\centerline{\includegraphics[clip,width=0.8\textwidth]{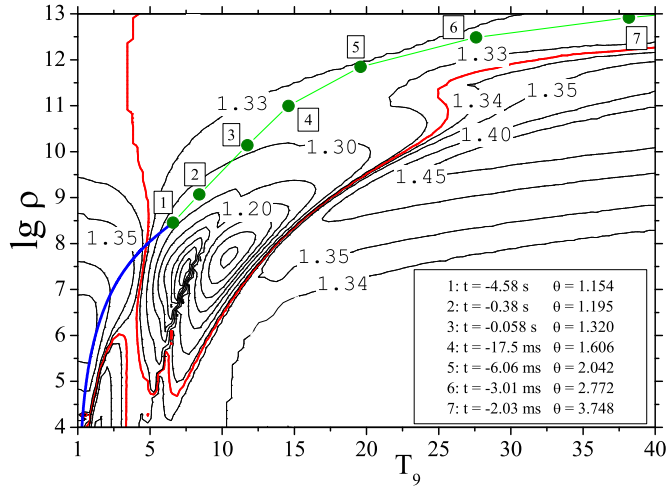}}
\caption{Track of a collapsing stellar core \cite{Yud2009} superimposed
on lines of constant adiabatic index $\gamma$ for $\theta = 30/26$.
}
\label{GamFeCollapse}
\end{figure}
Figure \ref{GamFeCollapse} shows an example of calculation of stellar
collapse with the NSE EOS described above.
Blue solid curve depicts
hydrostatic temperature-density structure of stellar core just
at the moment of loss of hydrostatic stability.
The solid line marked by black dots shows temporal evolution
of the collapsing core center.
The digits in squares nearby the dots are current number of line
in the inset table listing  times
left to the bounce  and central values of $\theta$ that varies
in time due to the process of neutronization.
The collapse starts at
time 4.58$\,$s before bounce when central temperature and density
are about \xmn{7}{9}K and \xmn{3}{8}g$\,$cm$^{-3}$.

\begin{figure}[htb!]
\centerline{\includegraphics[clip,width=0.8\textwidth]
{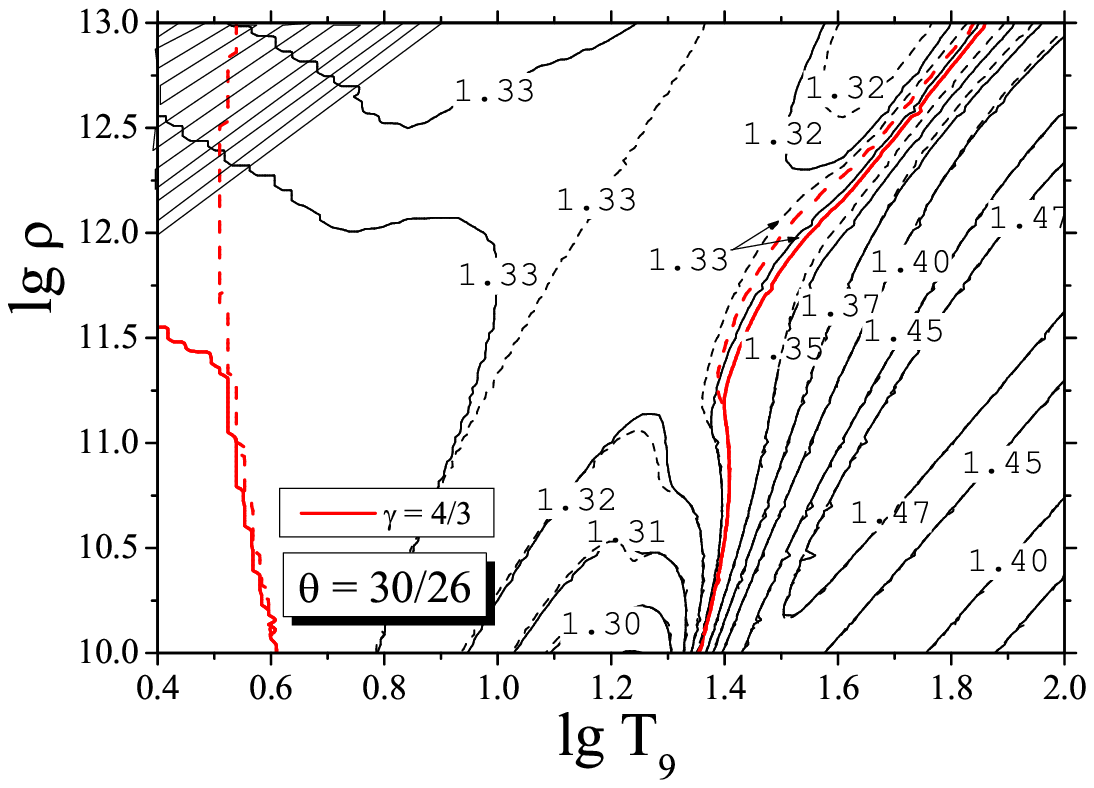}}
\caption{
}
\label{GamCoulombeps}
\end{figure}

The ideal gas approximation used in the NSE EOS above
should be examined for nonideal effects, especially for
the Coulomb and excluded volume interactions
(see papers \cite{NadYud05,Yudin2011} and references therein).

The Coulomb interaction between
numerous nuclides in a multi-component plasma,
such as shown in Fig. \ref{Yieps}
requires a detailed discussion \cite{NadYud05}.
The most significant contribution
to the NSE EOS comes from the ion–ion correlation
interaction. The typical values of the dimensionless
interaction parameter $\Gamma$ (which is equal in
order of magnitude to the ratio of the mean energy of
the interaction between two ions to their mean kinetic
energy) lie within the range 1--20 during the collapse.

Figure \ref{GamCoulombeps} shows
the result of calculations with taking into account the ion–ion
Coulomb interaction in the linear mixing model \cite{NadYud05}
which are indicated by solid lines.
The dashed lines indicate the results
of calculations without Coulomb interaction.
The critical value of $\gamma = 4/3$ is indicated by heavy
red lines. One can see that the Coulomb interaction makes
the instability ravine wider and deeper.
However, this effect becomes noticeable only at high
densities $\rho \gtrsim 10^{11}$g$\,$cm$^{-3}$ and
practically it does not change the after-bounce trajectory
of accreting shock wave shown in Fig. \ref{EncMass}.

\section*{Phase transitions in dense matter}

In recent years, substantial progress has been
achieved in solving the problem of supernova outburst
triggered by a {\em spherically symmetric\/} collapse
of stellar core. This problem has remained unsolved
for several decades.
However, the presence of initially strong
magnetic field and (or) rotation seems to be
sufficient to explain the supernova outburst,
as, for example, in the magneto-rotational
model of a supernova explosion
(see paper \cite{Bisnovaty08} and references therein)
and in the model of rotational fragmentation of
the collapsing core of a massive star (without any
magnetic field) into a close pair of neutron stars
(see paper \cite{Imshennik10} and references therein).

First, it has been shown (see, e.g., papers
\cite{Blondin05,ImshLitvin06,MarekJanka09} and references therein)
that the front of the quasi-steady-state
accretion shock separating the collapsed stellar core
from the stellar shell may become unstable
against three-dimensional perturbations, whose
growth gives rise to large-scale circulation mass
flows. As a result, an input of additional thermal
energy to the accretion shock could help to transform
it into a diverging blast wave capable finally
to eject the supernova envelope.

Second, when the central density of the collapsing core
becomes comparable or exceeds the nuclear one,
the phase transition from separate nuclei to nuclear matter
or from nuclear matter to quark one can occur.
In this case, at a sufficiently large density
jump on the interface between the two phases, the stellar core
loses its stability and undergoes additional contraction
in the hydrodynamic regime, forming a new shock during
its subsequent deceleration. This shock propagates outward and,
merging with the accretion shock, transforms it into a
diverging shock that triggers a supernova outburst
(see papers \cite{Sagert09,Fischer11,Krivoruchenko11,YudinRazNad13}
and references therein).
It was shown \cite{Sagert09,Fischer11} that an additional narrow
electron antineutrino peak appears in the neutrino light curve
as a specific signature of the phase transition to quark matter.
For a Galactic core-collapse supernova,
such a peak could be resolvable by the present neutrino
detectors.
 At present, there is extensive literature
on phase transitions in nuclear and, particularly,
quark matter. The physical properties of nuclear and
quark matter and their influence on the structure and
thermal evolution of super dense stars are described
in detail in the monograph \cite{Haense07}.

\section*{Conclusion}
The bulk of energy, that neutrino takes away from collapsing stellar
core, is radiated under the conditions of nuclear statistical
equilibrium (NSE) in the neutrino-opaque regime. In this case,
the neutrino heat conduction theory (NHC) is the best tool for modeling
the neutrino hydrodynamic processes. Contrary to the frequently used
approach based on the direct numerical solution of the neutrino transfer
equation in a comoving frame, the NHC allows
to detach the most CP-time consuming calculations of local functions
(the NSE matter equation of state and 3 kinetic coefficients
$A_\nu, B_\nu, C_\nu$
in equations of diffusion of the neutrino energy and leptonic charge)
from the difference scheme elaborating partial derivatives.
The local functions depend only on three arguments $(\rho , T, \theta)$.
One can compile in advance (only once!) detailed 3-entry tables for
the free energy $F(\rho , T, \theta)$ and coefficients $A_\nu, B_\nu, C_\nu$
and then use them to solve different neutrino hydrodynamic problem.
Of course, the tables should be supplemented with an algorithm of
interpolation. Special attention should be devoted to interpolation
of the free energy $F$ as a thermodynamic potential
generating pressure, specific energy, and entropy.
The interpolation algorithm should not violate
the continuity of $F$ and its first and second partial derivatives
by $(\rho , T, \theta)$. The most appropriate for this requirement
is the algorithm of {\em local\/} splines \cite{Ryaben2002}
that was successfully tested in Thesis \cite{Yud2009}.
The gravitational collapse of stellar cores has a remarkable
property. Different shells of the stellar core as being involved
into the collapse follow  the trajectories in space
$(\rho , T, \theta)$ close to that shown in Fig. \ref{GamFeCollapse}
for the stellar center. Thus, all stelar matter falls onto the center
moving along rather a narrow ``tube'' in $(\rho , T, \theta)$ space.
This allows to work with tables of a moderate size.

We showed that such physical constituents of the NSE equation of state
as nuclear excited states and coulomb and excluded volume interactions
are of a minor influence on hydrodynamics of the collapse.
However, they modify the composition of NSE matter and thereby can
change the rates of the neutrino-nuclear interactions which enter
the kinetic coefficients $A_\nu, B_\nu, C_\nu$.
This effect needs further detailed investigation.

The phase transitions possible in dense nuclear matter seem to be
of great importance for solving the long standing problem of supernova
outburst in case of spherically symmetrical stars without rotation
and magnetic fields.

\vspace*{3mm}
 {\bf Acknowledgement}.
  We are thankful to Prof. Remo Ruffini for his invitation
  to attend dedicated to the 70-th birthday of Prof. David Arnett
  ICRANet workshop ``From Nuclei to White Dwarfs and Neutron Stars'',
  (Les Houches, 3--8 April 2011).

   The work was supported
  by the Russian government grant 11.G34.31.0047,
  the Russian Foundation Basic Research (RFBR) grants
  11-02-00882-a, \mbox{12-02-00955-a},
  and the Swiss National Science Foundation SCOPES project
  No.~IZ73Z0-128180/1.

\end{document}